\documentclass{PoS}

\title{Pad\'e approximants and $g-2$ for the muon}

\ShortTitle{Pad\'e approximants for $g-2$}

\author{Christopher Aubin\\
        Department of Physics and Engineering Physics, Fordham University\\ Bronx, NY 10458, USA\\
}
        
\author{Thomas Blum\\
        Physics Department, 
University of Connecticut\\ Storrs, CT 06269, USA\\
}

\author{\speaker{Maarten Golterman}\thanks{Permanent address: Department of Physics and Astronomy,
San Francisco State University, San Francisco, CA 94132, USA.}\\
        Institut de F\'\i sica d'Altes Energies, Universitat Aut\`onoma de Barcelona,\\ E-08193 Bellaterra, Barcelona, Spain\\
}

\author{Santiago Peris\\
        Department of Physics, Universitat Aut\`onoma de Barcelona\\ E-08193 Bellaterra, Barcelona, Spain\\
}

\abstract{The leading hadronic contribution to the muon anomalous magnetic moment is given by a weighted euclidean momentum integral of the hadronic vacuum polarization. This integral is dominated by momenta of order the muon mass. Since  in lattice QCD  it is difficult to compute the vacuum polarization at a large number of low momenta, a parametrization of the vacuum polarization is required to extrapolate the data.  Most fits to date are based on vector meson dominance, which introduces model dependence into the lattice computation of the magnetic moment. Here we introduce a model-independent extrapolation method, and present a few first tests of this new method.}

\FullConference{The 30 International Symposium on Lattice Field Theory - Lattice 2012,\\
		June 24-29, 2012\\
		Cairns, Australia}

\begin{document}

\section{Introduction}
The anomalous magnetic moment of the muon $g-2$ has been measured with
great accuracy \cite{BNL}, and will be measured with even greater accuracy
in the near future.   Therefore, a reliable computation of $a_\mu=g-2$ from
theory with a comparable error would provide a precision test of the Standard
Model that is sensitive to a large class of models of new physics beyond the
Standard Model.   For this reason there has recently been a lot of interest in 
lattice computations of $a_\mu$ with controlled errors \cite{AB,FJPR2011,BDKZ2011,DJJW2012}; for an overview, and
more references, we refer to Ref.~\cite{TBlat12}.  Here, we report on recent work
on the leading hadronic contribution to $a_\mu$, which comes from the
hadronic vacuum polarization \cite{ABGS}.

\begin{figure}
\includegraphics[width=.48\textwidth]{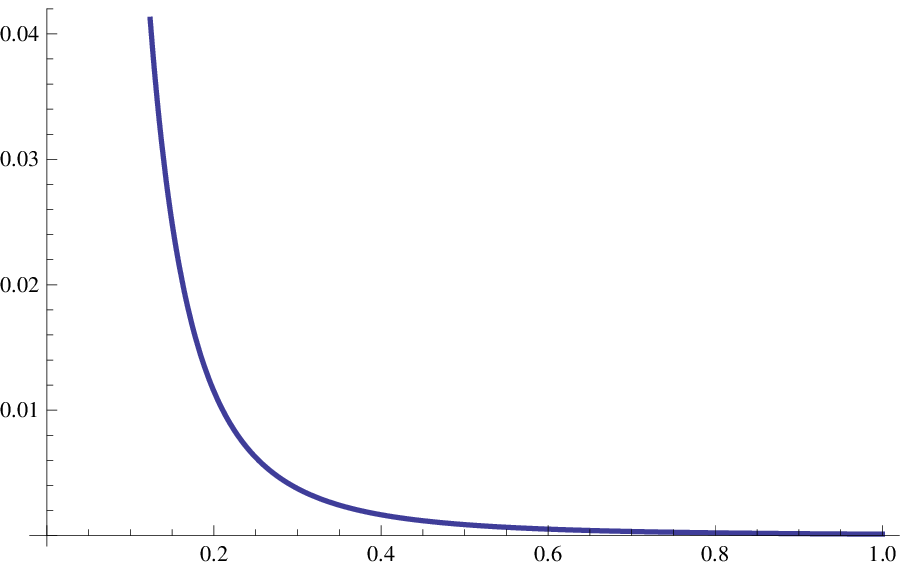}
\hspace{0.45cm}
\includegraphics[width=.48\textwidth]{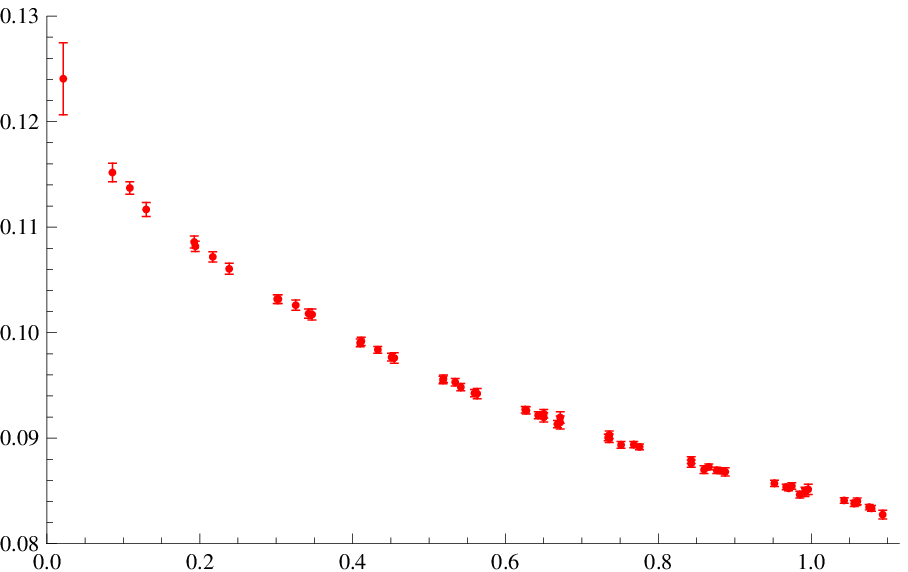}
\caption{Left panel: the weight $f(Q^2)$ for the muon;
right panel: typical data for $\Pi(Q^2)$ from the lattice.
Horizontal axis: $Q^2$ in GeV$^2$.}
\label{f}
\end{figure}

The contribution to $a_\mu$ from the lowest-order hadronic vacuum
polarization can be written as an integral over the subtracted vacuum
polarization $\Pi(Q^2)-\Pi(0)$ as a function of euclidean $Q^2$ \cite{LPdR,TB},
\begin{equation}
\label{amu}
a_\mu^{\rm HLO}=4\alpha^2\int_0^\infty dQ^2 f(Q^2)\left(\Pi(0)-\Pi(Q^2)\right)\ ,
\end{equation}
where $f(Q^2)$ is a kinematic weight shown in Fig.~\ref{f}, left panel.
The right panel shows a typical example of lattice data for $\Pi(Q^2)$
(these are the data from a $64^3\times 144$ lattice with lattice spacing
$0.06$~fm and $m_\pi=220$~MeV discussed below).  

Clearly, one needs to fit these data in order to compute the integral.
In most lattice computations of $a_\mu^{\rm HLO}$ to date this has been
done with various variants of vector-meson dominance (VMD).\footnote{
Ref.~\cite{DJJW2012} used Pad\'e approximants, but, as we will see below,
of a different type than those supported by a convergence theorem.}   This
introduces model dependence into the computation, and the aim of the
work presented here is to remove this model dependence.

\section{Multi-point Pad\'e approximants}
We start from the observation that we can write a subtracted dispersion
relation for $\Pi(Q^2)$:
\begin{equation}
\label{disprel}
\left(\Pi(0)-\Pi(Q^2)\right)/Q^2=\int_{4m_\pi^2}^\infty dt\;
\frac{\rho(t)}{t(t+Q^2)}\equiv\Phi(Q^2)\ ,
\end{equation}
in which $\rho(t)={\rm Im}\,\Pi(t)/\pi$ is the spectral function.   Because
the spectral function $\rho(t)\ge 0$, the integral in Eq.~(\ref{disprel})
is a Stieltjes function, analytic everywhere except along the cut
$(-\infty,-4m_\pi^2]$.

For such a function, there exists a theorem, proven in Refs.~\cite{B69,B73}:\\

\noindent{\bf Theorem:}
Given $P$ points $(Q_i^2,\Phi(Q_i^2))$, $i\in\{1,\dots,P\}$,
a sequence of Pad\'e approximants
(PAs) can be constructed which converge to $\Phi(Q^2)$ on any closed,
bounded region of the complex plane excluding the cut, in the limit $P\to\infty$.\\

This sequence of PAs can be constructed from the $P$ points through a
continued fraction:
\begin{equation}
\label{frac}
\Phi(Q^2)=\frac{\Phi(Q_1^2)}{1+\mbox{\Large{${\frac{(Q^2-Q_1^2)\Psi_1(Q_2^2)}
{1+\ { }_{{\ddots}_{{\ \frac{(Q^2-Q_{P-1}^2)\Psi_{P-1}(Q_P^2)}{1+(Q^2-Q_P^2)\Psi_P(Q^2)}}}}}}$}}}\ ,
\end{equation}
with $\Psi_i$ related to $\Phi(Q_{j\le i+1}^2)$ ($\Psi_0=\Phi(Q_1^2)$, {\it etc.}).
Equation~(\ref{frac}) yields a $[[(P-1)/2],[P/2]]$ PA (where $[x]$ is the integer
part of $x$).   Furthermore, one can prove that this can be
rewritten as \cite{B69,B73,BG1996}
\begin{equation}
\label{poles}
\Pi(Q^2)=\Pi(0)-Q^2\left(a_0+\sum_{n=1}^{[P/2]}\frac{a_n}{b_n+Q^2}\right)\ ,
\end{equation}
with 
\begin{equation}
\label{conditions}
a_n>0\ ,\qquad b_{[P/2]}>\dots>b_1>4m_\pi^2\ ,
\end{equation}
{\it i.e.}, all poles are single poles, they are located on the cut, and all residues are positive.   The
constant $a_0=0$ for $P$ even.

In the situation of an actual fit to data for $\Pi(Q^2)$ obtained from a numerical computation, these data are only known within some statistical errors. That implies that we do not know any points of the function exactly, and a multi-point sequence of PAs as
implied by the theorem cannot be constructed.  Our strategy will be to fit a fixed number of data points on a given interval, using the fact that since $\Pi(Q^2)$, according to the theorem, can be described by a converging sequence of PAs of
the form~(\ref{poles}), this equation provides a valid functional form to which to fit the data.   More concretely, we will fit the form~(\ref{poles}) for
$P\in\{2,3,4,5\}$; this yields $[0,1]$, $[1,1]$, $[1,2]$ and $[2,2]$ PAs.
In order to compare diffferent fits, we will then compute
\begin{equation}
\label{amutest}
a_\mu^{\rm HLO,Q^2\le 1}=4\alpha^2\int_0^{1~{\rm GeV}^2} dQ^2 f(Q^2)\left(\Pi(0)-\Pi(Q^2)\right)\ .
\end{equation}
We note that VMD is the same as a $[0,1]$ PA, but keeping $b_1=m_\rho^2$
fixed:  This is {\it not} a valid PA in the sense of the theorem, because the theorem
does not say anything about the possible values of the parameters in 
addition to the conditions~(\ref{conditions}).

\section{Tests}
For our first test, we explore fits to a MILC data set on a $28^3 \times 96$
lattice with lattice spacing $0.09$~fm, and a pion mass $m_\pi^2\approx
480$~MeV \cite{MILC}.   This is one of the data sets that was also used in Ref.~\cite{AB}.
We show the results for the $P\in\{2,3,4,5\}$ PAs and for VMD in Table~\ref{t1}.  The uncorrelated VMD fit is the same as the fit to these data performed
in Ref.~\cite{AB}, and the results agree.

\begin{table}
\begin{center}
\begin{tabular}{|c|c||l|c||c|c|}
\hline
 \multicolumn{2}{|c||}{} 
& \multicolumn{2}{c||}{\bf correlated} & \multicolumn{2}{c|}{\bf uncorrelated}
\\
 \multicolumn{2}{|c||}{}  & \multicolumn{2}{c||}{interval $0<Q^2\le 0.6$~GeV$^2$} & \multicolumn{2}{c|}{interval $0<Q^2\le 1$~GeV$^2$}
\\
\hline\hline
PA & \# parameters & $\chi^2$/dof & $10^{10}a_\mu^{{\rm HLO},Q^2\le 1}$
& $\chi^2$/dof & $10^{10}a_\mu^{{\rm HLO},Q^2\le 1}$ \\
\hline
VMD & 2 & 5.86/3$^*$ & 363(7) &  4.37/18 & 413(8)   \\
$[0,1]$ & 3 & 11.4/8 & 338(6) & 3.58/17 & 373(37)    \\
$[1,1]$ & 4 & 7.49/7 & 350(8) & 3.36/16 & 424(116)   \\
$[1,2]$ & 5 & 7.49/6 & 350(8) & 3.35/15 & 443(293)  \\
$[2,2]$ & 6 & 7.49/5 & 350(7) & 3.35/14 & 445(432)  \\
\hline
\end{tabular}
\caption{
Results for various fits.  The fit marked 
\null$^*$ was done on an interval $0<Q^2\le 0.35$~GeV$^2$.}
\label{t1}
\end{center}
\end{table}%

Table~\ref{t1} leads us to make the following observations:
\begin{itemize}
\item{} The correlated VMD fit is a bad fit as measured by $\chi^2$ per
degree of freedom (dof); adding parameters the fits clearly improve.
Note that we always choose the fitting interval by looking for a minimal
value of $\chi^2$/dof.
\item{}  It turns out that it is difficult to determine the parameters of the
second pole with any precision \cite{ABGS} (as can be inferred from the
values of $\chi^2$/dof), but $a_\mu^{\rm HLO,Q^2\le 1}$
is insensitive to the second and higher poles.
\item{}  There is good internal consistency between all fits shown in the
table, except between the uncorrelated VMD fit and any of the correlated
PA fits.   However, the VMD fits are model dependent, which translates 
into an unknown systematic error in these fits.
\end{itemize}

\begin{figure}
\hspace{-0.1cm}
\includegraphics[width=.55\textwidth]{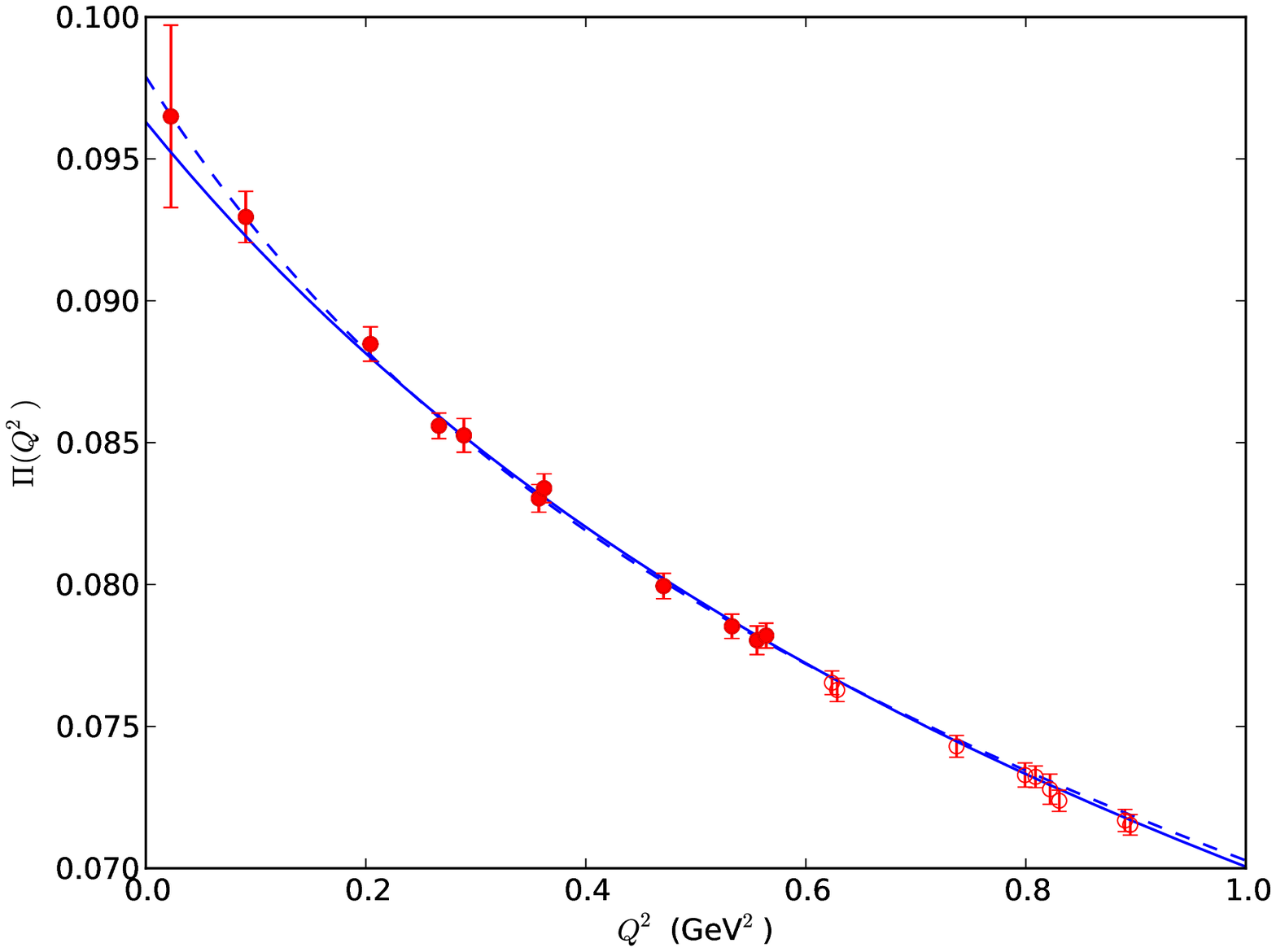}
\hspace{-0.9cm}
\includegraphics[width=.55\textwidth]{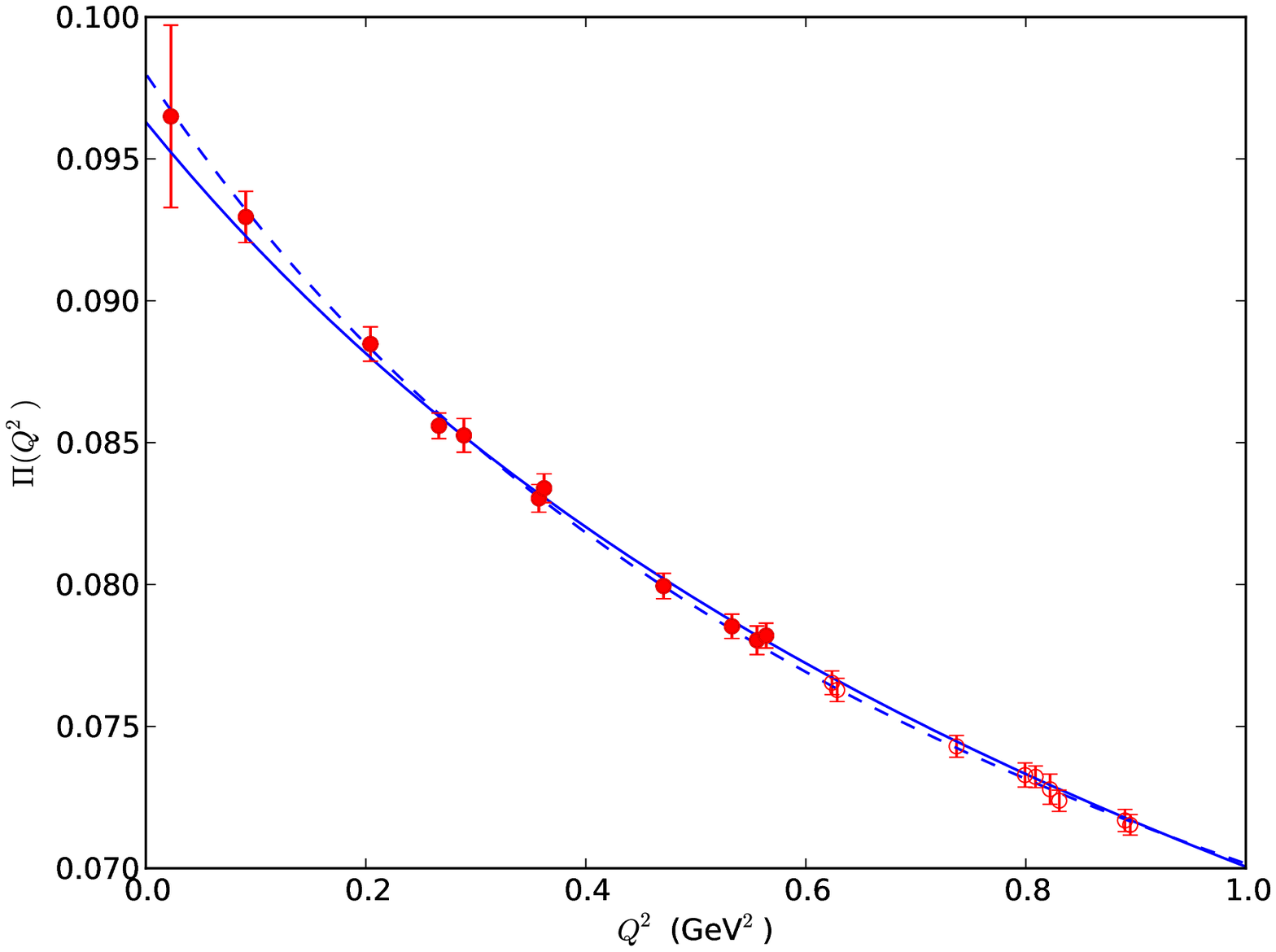}
\caption{Correlated (solid curve) and uncorrelated (dashed curve)
for the $[1,1]$ PA fits (left panel), and for the correlated $[1,1]$ (solid curve) and uncorrlated VMD (dashed curve) fits (right panel).}
\label{fits}
\end{figure}

We display some of the fits of Table~\ref{t1} in Fig.~\ref{fits}.    Not
surprisingly, the uncorrelated fits look better at small $Q^2$, but all
fits shown in the figure do a good job of describing the data.   Therefore, based on the
data, it is not possible to decide which of these fits is the best fit.   

We repeated our explorative analysis on MILC lattices on a $64^3\times 144$
lattice with lattice spacing $0.06$~fm, and $m_\pi\approx 220$~MeV.
We find very similar results;\footnote{Of course, central values of
$a_\mu^{\rm HLO,Q^2\le 1}$ are quite different, if only because of the
smaller pion mass.} in particular we find
\begin{eqnarray}
\label{superfine}
a_\mu^{\rm HLO,Q^2\le 1}&=&572(41)\times 10^{-10}\ ,\qquad [1,1]\ \mbox{correlated}\ ,\\
a_\mu^{\rm HLO,Q^2\le 1}&=&646(8)\times 10^{-10}\ ,\qquad
\mbox{VMD\ uncorrelated}\ .\nonumber
\end{eqnarray}
Our conclusions are the same as before.   We note that for both data
sets the discrepancy between the correlated $[1,1]$ PA and the uncorrelated VMD fit is about
15\%.  {}From the point of view that both types of fit give a good description
of the data, we take this to imply that there is a systematic error of (at least)
this size afflicting the determination of $a_\mu^{\rm HLO}$ from the lattice.

\begin{figure}[t]
\begin{center}
\includegraphics[width=0.75\textwidth]{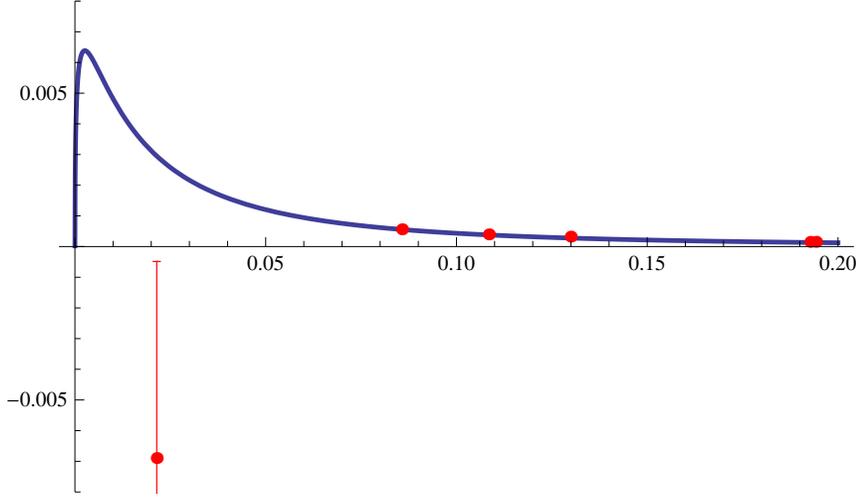}
\caption{The integrand of Eq.~(1.1), using the correlated $[1,1]$
PA fit to the $64^3\times 144$ data set (solid curve), compared with
the data weighted by $f(Q^2)$ in Eq.~(1.1).}
\label{integrand}
\end{center}
\end{figure}

The underlying problem is displayed in Fig.~\ref{integrand}, where we see
that there are essentially no data in the region dominating the integral
in Eq.~(\ref{amu}).   For this, one clearly needs data at more low values 
of $Q^2$, with smaller errors.   It would be interesting to see whether
these improvements can be attained by using twisted boundary conditions,
something that has been tried in this context in Ref.~\cite{DJJW2012}, and by an
error reduction technique such as that proposed in Ref.~\cite{BIS}.

We have also compared our PA fits with polynomial fits; results are shown 
in Table~\ref{t2}.   ``Poly~$n$'' indicates a fit with a polynomial of degree $n$.
All fits are correlated fits; and the pairs of fits ``Poly~3,'' ``$[1,1]$'' respectively
``Poly~4,'' ``$[1,2]$'' have the same number of parameters.   We observe
that the fits deteriorate in the polynomial case going from Poly~3 to Poly~4,
with errors increasing, and central values for $a_\mu^{{\rm HLO},Q^2\le 1}$
fluctuating more, while this is not the case going from the $[1,1]$ to the
$[1,2]$ PA fit.

\begin{table}
\begin{center}
\begin{tabular}{|c|c|c|c|c|c|c|c|c|}
\hline
& \multicolumn{2}{c|}{Poly~3} & \multicolumn{2}{c|}{Poly~4}
& \multicolumn{2}{c|}{PA~[1,1]} & \multicolumn{2}{c|}{PA~[1,2]}\\
\hline\hline
\# points & $\chi^2$/dof & $a_\mu^{(1)}$
& $\chi^2$/dof & $a_\mu^{(1)}$
& $\chi^2$/dof & $a_\mu^{(1)}$
& $\chi^2$/dof & $a_\mu^{(1)}$\\
\hline
16 & 9.6/12 & 543(35) &  9.5/11 & 483(244) & 9.7/12 & 564(55) & 9.7/11 & 565(41)  \\
18 & 11.4/14 & 526(33) & 10.5/13 & 596(79) & 11.2/14 & 541(46) & 11.5/13 & 561(21)   \\
20 & 13.1/16 & 536(23) & 13.1/15 & 535(45) & 13.9/16 & 572(41) & 13.9/15 & 572(37)  \\
22 & 16.5/18 & 541(23) & 15.9/17 & 513(44) & 18.5/18 & 566(37) & 18.5/17 & 566(33) \\
24 & 16.6/20 & 537(18) & 16.4/19 & 521(41) & 19.4/20 & 583(34) & 19.4/19 & 583(33) \\
26 & 30.7/22 & 505(16) & 23.6/21 & 580(32) & 26.8/22 & 557(31) & 26.7/21 & 560(27)  \\
\hline
\end{tabular}
\caption{Comparison of polynomial with PA fits, abbreviating
$a_\mu^{{\rm HLO},Q^2\le 1}$ by $a_\mu^{(1)}$.  The number of data points
included in the fit is indicated in the first column, with
20 points corresponding to the fitting interval $0<Q^2\le 0.53~{\rm GeV}^2$.
Data from the $64^3\times 144$ MILC lattices.}
\end{center}
\label{t2}
\end{table}%

\section{Conclusions}
We presented a new method for parametrizing the momentum dependence of the hadronic vacuum polarization, with the aim to avoid the model dependence of the VMD-based
fits that up to this point have been used in most fits to lattice data for the
hadronic vacuum polarization.   It turns out that this is possible, because
the vacuum polarization can be represented in terms of a Stieltjes function,
for which sequences of Pad\'e approximants can be constructed which
converge uniformly to the function on any bounded region in the complex
$Q^2$ place excluding the cut.

We have tested this new idea on two examples of lattice data
for the vacuum polarization.   We note that the fits based on 
Pad\'e approximants can lead to larger statistical errors than some of 
the VMD fits, as for instance in Eq.~(\ref{superfine}).   However, it should
be emphasized that the latter are afflicted with an unknown systematic
error originating in the inherent model dependence of VMD-based fits.
The fits based on Pad\'e approximants avoid this systematic error.\footnote{Of
course, there are other systematic errors, such as scaling violations,
finite-size effects, and chiral extrapolation errors.}  

The new method looks
promising.   However, it is clear that data for the hadronic vacuum polarization
at more low $Q^2$ values (of order the square of the muon mass), and with
smaller errors, will be needed in order to reach a higher precision for 
$a_\mu^{\rm HLO}$.   As we have seen, fits based on Pad\'e approximants and VMD-based fits (both correlated and uncorrelated) give a good description
of the data, but lead to values for $a_\mu^{{\rm HLO},Q^2\le 1}$ which differ
by about 15\%.   

Finally, we observe that $g-2$ is an example of a quantity
which is quite sensitive to the value of the pion mass.   Therefore,
better data for the hadronic vacuum polarization
will also have to be obtained at small values of the pion mass, certainly
significantly smaller than $300$~MeV.   

\vspace{1cm}
\noindent
{\bf Acknowledgements}
We would like to thank USQCD for the computing resources used to generate the vacuum polarization as well as the MILC collaboration for providing the configurations used.
TB and MG are supported in part by the US Department of Energy under
Grant No. DE-FG02-92ER40716 and Grant No. DE-FG03-92ER40711.
MG is also supported in part by the Spanish Ministerio de Educaci\'on, Cultura y Deporte, under program SAB2011-0074.
SP is supported by CICYTFEDER-FPA2008-01430, FPA2011-25948, SGR2009-894,
the Spanish Consolider-Ingenio 2010 Program
CPAN (CSD2007-00042) and also by the Programa de Movilidad
PR2010-0284.

\end{document}